\documentclass[aps,showpacs,prl,twocolumn,groupedaddress]{revtex4-1}

\usepackage{amssymb}
\usepackage{amsmath}
\usepackage{graphicx}
\usepackage{dcolumn}
\usepackage{bm}

\begin{document}


\title{Emergence of hierarchical networks and polysynchronous behaviour in simple adaptive systems}


\author{V. Botella-Soler}
\email{vicente.botella@uv.es} \affiliation{Departament de
F\'isica Te\`orica and IFIC, Universitat de Val\`encia-CSIC,
46100-Burjassot, Val\`encia, Spain}
\author{Paul Glendinning}
\email{p.a.glendinning@manchester.ac.uk}
 \affiliation{School of Mathematics and\\
 Centre for Interdisciplinary Computational and Dynamical Analysis (CICADA),\\
 University of Manchester, Manchester M13 9PL, U.K.}



\begin{abstract}
We describe the dynamics of a simple adaptive network. The
network architecture evolves to a number of disconnected
components on which the dynamics is characterized by the
possibility of differently synchronized nodes within the same
network (polysynchronous states). These systems may have
implications for the evolutionary emergence of polysynchrony
and hierarchical networks in physical or biological systems
modeled by adaptive networks.
\end{abstract}


\maketitle


Graphs or networks are used to describe the evolution and
interaction of coupled systems. Each node is assumed to have an
underlying dynamics which can be affected by `neighbouring'
nodes. This influence is represented by a directed edge, or
arrow, indicating input from one node to another in the
direction of the arrow. This can be complicated by allowing the
neighbours of a node to vary with time as a function of the
state of the nodes, giving rise to \emph{adaptive networks}
\cite{Gross2008, gross2009adaptive}. These models have been
applied to a broad range of problems including the connectivity
of the internet, the social interactions in a community and
motifs in systems biology, see \cite{boccaletti2006complex,
milo2002network, Gross2008, gross2009adaptive} for further
details and examples.

The dynamics of these systems can be rich, both in terms of the
temporal dynamics at the nodes and the emergence of topological
structure in the networks. For example, the dynamics can
synchronize, so all nodes behave asymptotically in the same way
\cite{pikovsky2003synchronization}. Polysynchrony describes a
more subtle form of synchronization
\cite{stewart2003symmetry,golubitsky2004some,field2004combinatorial,
aguiar2009dynamics, agarwal2010dynamical} where groups of nodes
can be synchronized and, unlike cluster synchronization,
synchronized nodes need not be connected to each other.

The aim of this note is to show that complicated
polysynchronous dynamics can emerge in adaptive networks using
a simple homophilic principle to determine the evolution of the
links of the network. This is based on the idea that nodes
`like' being connected to similar nodes, a common assumption in
many socially motivated networks and a basic principle in
neural network dynamics. In other words we consider a
collection of identical individuals operating under the same
homophilic rules, and show that the system evolves to a finite
number of connected components, each of which can display
polysynchronous states. All the examples we have looked at have
a further interesting property: the final states have graphs
which have a hierarchical structure, with a group of strongly
connected components at the bottom, and then the remaining
nodes being arranged above these (taking information from them,
but not giving information) in levels reflecting the number of
directed links needed to get to the node from one of the bottom
nodes or `roots'. (Note that we say a set of nodes is \emph{strongly
connected} if there is a path in the graph following the directed
edges or arrows between any two nodes, and \emph{connected} if there
is a path between any two nodes following edges, so the path can use an edge
in the opposite direction to the arrow.) This suggests that polysynchrony and
hierarchy are natural states that can emerge from simple
undifferentiated networks over time. We will discuss further
implications of this towards the end of the article.

\begin{figure}[h]
\centerline{
\includegraphics[width=6cm]{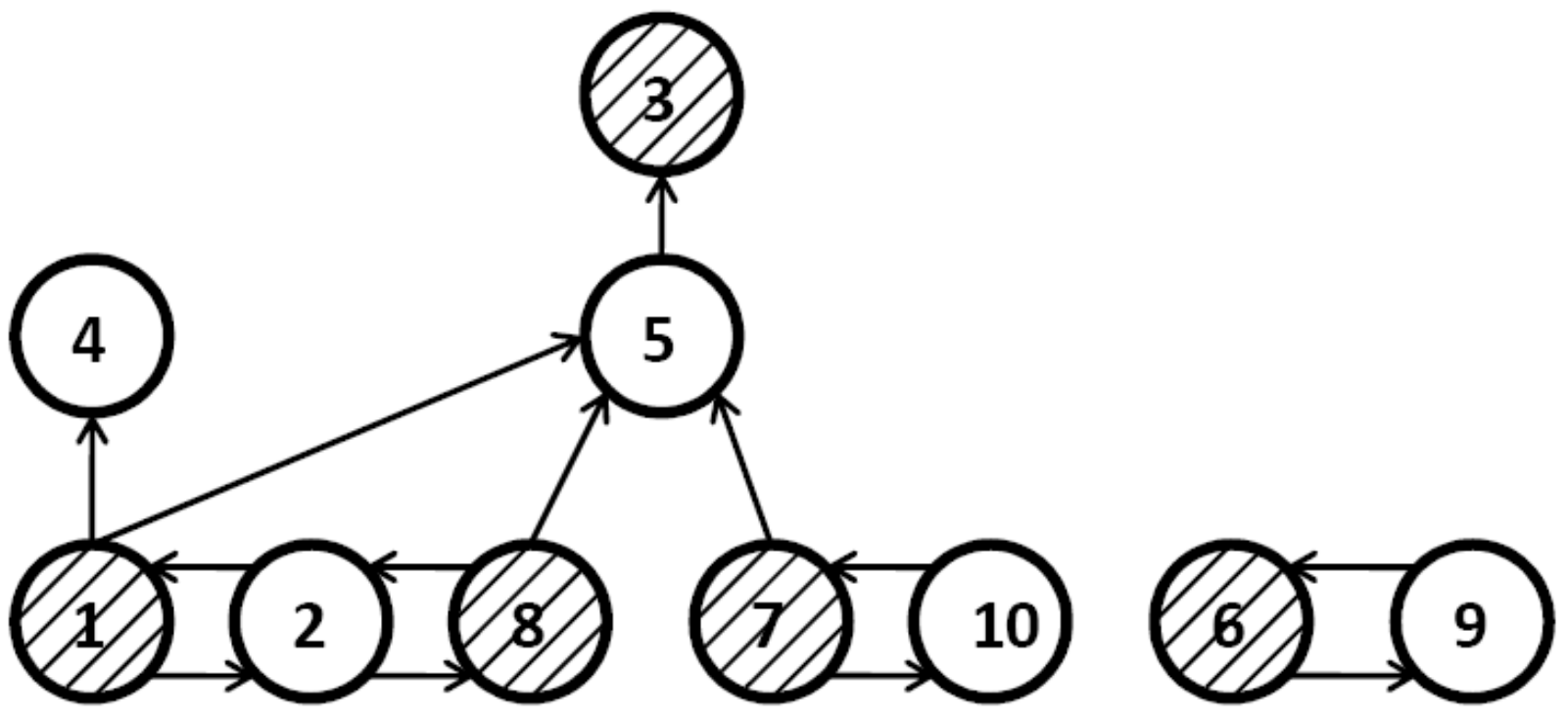}
} \caption{Example of a polysynchrony pattern ($N=10$,
$\varepsilon=0.85$). In this case the node dynamics have been
attracted to two different fixed points. Nodes filled with the
same pattern are synchronized including the isolated pair
$\{6,9\}$. Arrows have an associated weight and equation
(\ref{eq:mneigh}) tells us that, for each node, the weights of
all incoming arrows sum up to $m=N-1$. Since a node always get
its inputs from only one synchrony class, we have avoided
labelling the weights for the sake of clarity.} \label{fig:one}
\end{figure}
\begin{figure}[h]
\centerline{
\includegraphics[width=8cm]{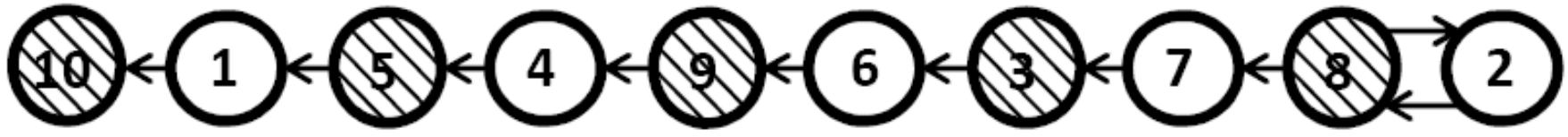}
} \caption{Example of a polysynchrony pattern ($N=10$,
$\varepsilon=0.8$). The node dynamics are chaotic in this case.
Nodes filled with the same pattern are synchronized.}
\label{fig:two}
\end{figure}

Before describing our model in full detail it is worth looking
at a couple of examples to illustrate the results.
The first (Fig.~\ref{fig:one}) shows the final state of
our system below with ten nodes. The final graph has broken
into two subsystems; one containing two nodes and the other
eight nodes. The eight node subsystem has two basic strongly
connected sets, $\{1,2,8\}$ and $\{7,10\}$, and these influence
but are not influenced by nodes 4 and 5, which in turn
influences node 3. There are two polysynchrony classes: the
dynamics of nodes $\{2,4,5,9,10\}$ is identical as is that of
the remaining nodes. At first sight this appears remarkable as
the nodes in the isolated pair $\{6, 9\}$ are also part of the
same polysynchrony classes. We explain this below. In this case
the dynamics is trivial: two fixed points, one for each
polysynchrony class, but more complicated behaviour is possible
as shown in Fig.~\ref{fig:two}.

The final state in Fig.~\ref{fig:two} is a chain based on two
interacting nodes at the root with polysynchrony classes that
oscillate up the chain. The dynamics of each polysynchrony
class is chaotic, but each element of the same class is fully
synchronized.

To summarize from these two observations: the dynamics breaks
up into a small number of disconnected sets, and within each of
these sets the dynamics may be either simple (a fixed point in
the first example) or more complicated (chaotic in the second
example), and in fact we will show that different behaviours
across polysynchrony classes in the same connected component
are also possible. Contrary to full synchronization or cluster
synchronization, the synchronized nodes have no direct links
connecting them. Moreover the final graphs have a hierarchical
structure, lending themselves to interpretation as motifs as
they allow a well-defined sense of input and output nodes.

The model consists of a directed network of $N$ nodes. The
dynamics of the $i$th ($i=1,\dots,N$) node are given by
\begin{equation}
x^{i}_{n+1}=f(x^{i}_{n})+\frac{\varepsilon}{m}\sum_{j=1}^{N}A^{ij}_{n}(f(x^{j}_{n})-f(x^{i}_{n}))
\label{eq:localdyn}
\end{equation}
where we choose $f$ to be the fully-chaotic logistic map
$f(x)=4x(1-x)$ and $A_n$ is the adjacency matrix of the network
at time step $n$, so $A^{ij}_{n}=k$ if there are $k$
connections (directed edges) from $j$ to $i$, with the
direction indicated by the arrow in the accompanying figures.
Each node is assigned a fixed number $m$ of incoming links so
\begin{equation}
\sum_{j=1}^{N}A^{ij}_{n}=m, \label{eq:mneigh}
\end{equation}
and we choose $m=N-1$ throughout this paper. These links can be
rewired at each iteration following an algorithm that will be
described below.

At each iteration the $i$th node is influenced by the dynamics
of those nodes to which it is connected by an incoming arrow.
We will call these nodes the \textit{neighbours} of node $i$.
Due to the condition imposed by (\ref{eq:mneigh}), a node can
have at most $m$ neighbours.

We describe now the rewiring mechanism. At each iteration $n$
we compute the distance matrix $D^{ij}_n$
\begin{equation}
D^{ij}_n=\left\{
\begin{tabular}{ll}
      $|f(x^{i}_{n})-f(x^{j}_{n})|$, & \ \ $A^{ij}_n\neq0$ \\
      $0$, & \ \ $A^{ij}_n=0$
     \end{tabular}
     \right.
\end{equation}
and calculate from it the mean distance of a node to all its
neighbours
\begin{equation}
\langle
D\rangle^{i}_{n}=\frac{1}{a^{i}_{n}}\sum_{j=1}^{N}D^{ij}_{n}
\end{equation}
where $a^{i}_{n}$ is the unweighted number of neighbours of node $i$ at
time step $n$, i.e. the sum over $j$ of ${\rm sign}(A^{ij}_{n})$.

For the rewiring we apply a homophilic principle: nodes prefer
to be connected to nodes being in a similar state. We identify
the \textit{bad} neighbours $j\in\mathcal{B}^{i}_n$ of each
node $i$ at iteration $n$ with the following criterion
\begin{equation}
j\in\mathcal{B}^{i}_n \quad\text{if}\quad D^{ij}_{n}>\langle
D\rangle^{i}_{n},
\end{equation}
so a neighbour $j$ is considered \textit{bad} if its distance
$D^{ij}_{n}$ to the node is larger than the average distance of
the neighbourhood $\langle D\rangle^{i}_{n}$. The \textit{good}
neighbours of node $i$ are then given by
\begin{equation}
\mathcal{G}^{i}_n=\{1,\dots ,N\}\backslash\left(\mathcal{B}^{i}_n \cup\{i\}\right),
\end{equation}
so $i\notin \mathcal{G}^{i}_n$ which means that a node never becomes linked to
itself. Once the good and bad neighbours have been identified node $i$
will break the links coming from $\mathcal{B}^{i}_n$ and
randomly rewire them to nodes in $\mathcal{G}^{i}_n$. Let
$b^{i}_n$ be the number of \textit{bad} neighbours
\begin{equation}
b^{i}_n=\sum_{j\in\mathcal{B}^{i}_n}A^{ij}_{n}.
\end{equation}
Now choose $b^{i}_n$ elements of $\mathcal{G}^{i}_n$ at random
and suppose that $r^{ik}_n$ is the number of times node $k$ is
chosen. The adjacency matrix at the next time step is
\begin{equation}
A^{ik}_{n+1}=\left\{
\begin{tabular}{ll}
      $0$, & $k\in\mathcal{B}^{i}_n$ \\
      $A^{ik}_{n}+r^{ik}_n$, & $k\in\mathcal{G}^{i}_n$
     \end{tabular}
     \right..
\end{equation}

In all the cases described here the initial connectivity is
the symmetric all-to-all connectivity where each node in the network is
connected to all the possible $m=N-1$ neighbours and $A_0^{ii}=0$.

The first main observation is that, at least for systems of
moderate size, the rewiring dynamics always settles down to a
frozen state after a finite amount of time. This is possible
due to the criterion used to perform the rewiring of the links
at each iteration. If at time step $n$, a node $i$ has all its
incoming links coming from the same node or from a
synchronized set of neighbours, all elements in the distance
matrix $D^{ij}_n$ will have the same value ($D^{ij}_n=\langle
D\rangle^{i}_{n}$ for all $j$ such that $A^{ij}_n\neq 0$)
resulting in an empty set of \textit{bad} nodes and hence
`locking' node $i$ to its neighbourhood for the next time step.
However, we should notice the locking process is not
purely probabilistic because the interactions between the nodes
correlate their dynamics thus affecting the locking
probabilities.

\begin{figure}[h]
\centerline{
\includegraphics[width=8cm]{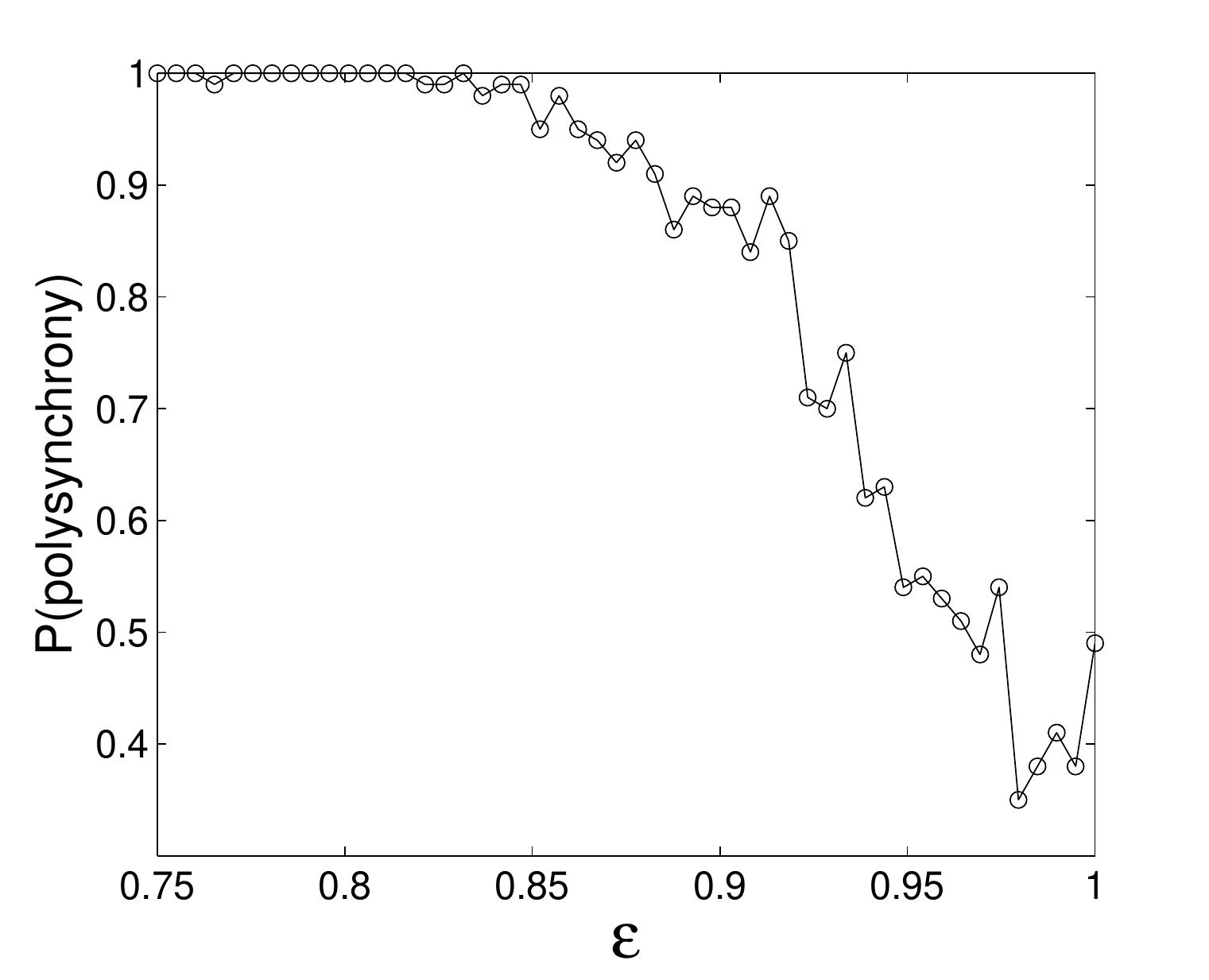}
} \caption{Probability of polysynchrony as a function of the
coupling strength $\varepsilon$. $N=10$, 100 initial conditions
have been evaluated for each value of $\varepsilon$.
Polysynchrony has been identified as explained in the text.}
\label{fig:polysynchrony}
\end{figure}

The second and most important observation is that, as
Fig.~\ref{fig:polysynchrony} shows, the dynamics of the nodes
once the network has frozen has a high probability of being
polysynchronous for a wide range of coupling strengths, namely
$0.75<\varepsilon<1$ and also
$0.12\lesssim\varepsilon\lesssim0.2$. There are other possible
dynamics for this adaptive network model depending on the value
of the coupling strength including uncorrelated dynamics (no
two nodes are synchronized), full synchronization (all the
nodes in the network synchronize) and cluster or partial
synchronization, but in the regions of parameter space that we
have identified, polysynchronous and hierarchical states appear
most frequently.

As studied in \cite{stewart2003symmetry} the network needs to
satisfy some symmetry requirements in order to achieve robust
polysynchrony. The structure of polysynchronous networks requires
a balanced equivalence relation to be established on the nodes,
so the nodes can be separated into equivalence classes of
nodes with equivalent inputs. For example, in
Fig.~\ref{fig:one} none of the nodes in the same synchrony
class have links connecting them to each other, but receive all
their $m$ inputs from nodes in the other synchrony class. In
principle, a more general input equivalence could be
established in which nodes in a certain synchrony class receive
inputs from various synchrony classes (including inputs from
their own synchrony class) \cite{golubitsky2005patterns}.
However, using arguments similar to those for the locking
mechanism we can show that this is not possible in our model
due to the homophilic rewiring criterion chosen: if a node
receives inputs from two different synchrony classes
a certain time step $n$ then typically the distance from these sets
will be different and all the links from one class will be rewired
at the next time step \cite{bg2011polyad}. We have taken advantage of this
property of the model to assess the polysynchronous character
of the final network by comparing the final difference
($\delta^{ij}_n=|x^i_n-x^j_n|$) and adjacency matrices. If all
the non-diagonal $0$ elements in the difference matrix (which
reflect synchrony between two nodes) correspond to
non-connected nodes (reflected by $0$ elements in the adjacency
matrix) then our network is polysynchronous.

\begin{figure}[h]
\centerline{
\includegraphics[width=8cm]{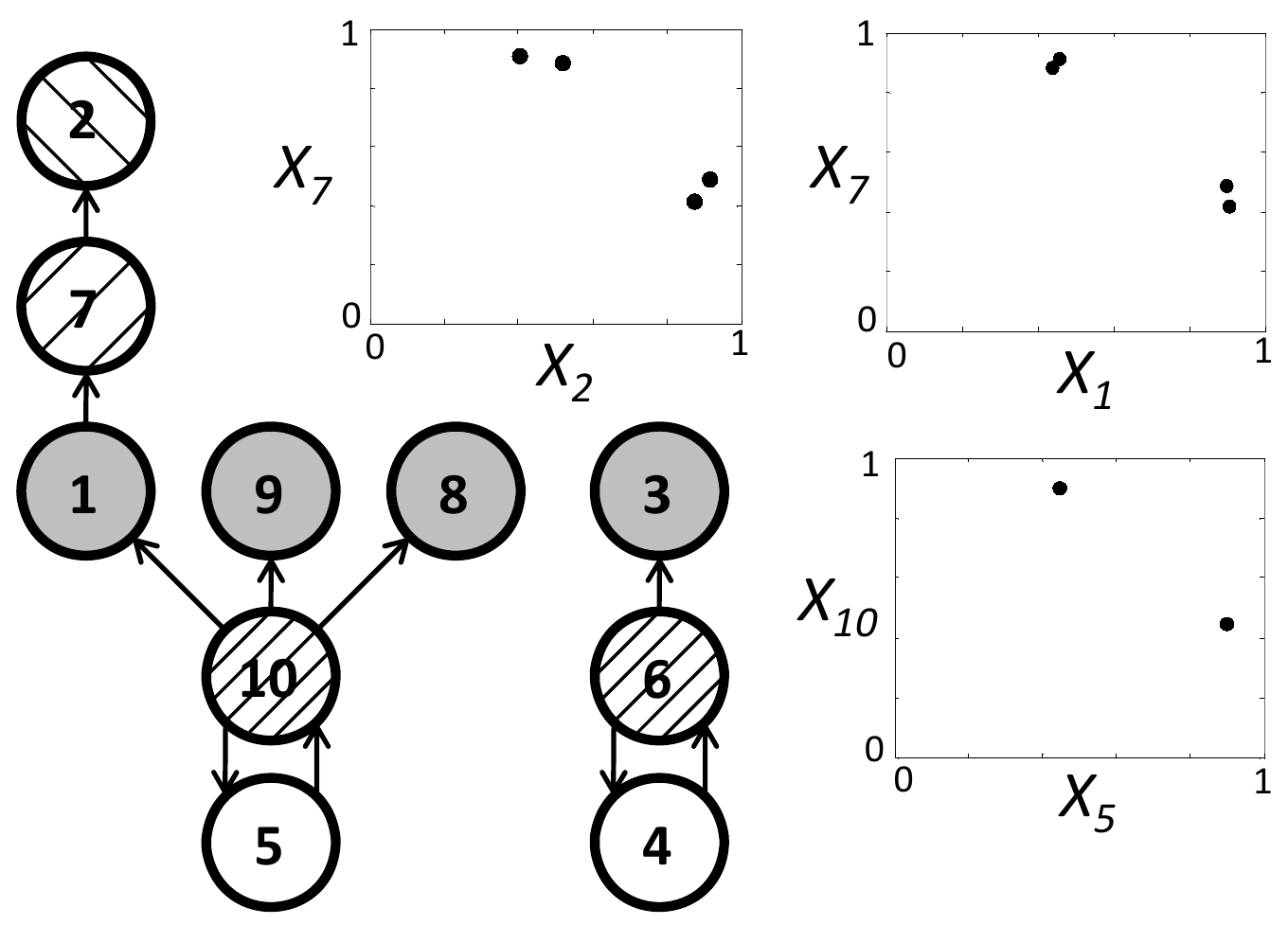}
} \caption{Example of a polysynchronous pattern ($N=10$,
$\varepsilon=0.14$). In this case there are 4 different
synchrony classes showing 2 different dynamical behaviours:
$\{4,5\}$ and $\{6,10\}$ are synchronized in a period-2 cycle
while $\{2\}$,$\{7\}$ and $\{1,3,8,9\}$ show period-4
dynamics.} \label{fig:four}
\end{figure}

A further characteristic of the networks showing polysynchrony
in our model is that their structure is usually strongly
hierarchical. If we define the set of nodes in strongly
connected components to be the `roots' of the network (for
instance, nodes 2 and 8 are the root of the network in
Fig.~\ref{fig:two}) then  the `level' of a node is the minimum
(directed) path length from the root to the node, and the
`height' of a node is the average distance over all possible
directed paths from the root to the node. These concepts are
inspired by the definitions of `trophic level' and `trophic
height' introduced for the study of food webs
\cite{quince2005topological}, and our choice of the root
ensures that the level and height are well-defined
\cite{bg2011polyad}. We say that a network is strongly
hierarchical if level and height coincide for all the nodes in
the network. One can see this definition applies to the
networks in Fig.~\ref{fig:one} and \ref{fig:two}. Another
example of strongly hierarchical polysynchronous network is
shown in Fig.~\ref{fig:four} where the root is composed of two
strongly connected pairs. This example has the additional
interest that different synchrony classes inside the same
cluster show different dynamical behaviour. While the root
pairs $\{5,10\}$ and $\{4,6\}$ follow a period-2 cycle, the
rest of the nodes in the network show period-4 dynamics.
Moreover, it is interesting to note that nodes in classes
\{4,5\} and \{1,3,8,9\} all receive the same input (from class
\{6,10\}) but show different dynamics. This illustrates that
having equivalent inputs is only a necessary but not sufficient
condition to be in the same synchrony class
\cite{golubitsky2006winding}.
Finally, note that since an element of a given polysynchrony
class receives its inputs from a given other synchrony class,
the equations determining the dynamics of each polysynchrony
node are the same. This implies that the after identifying all
nodes in the same polysynchrony class the dynamics is based on
a very small number of possibilities: we conjecture that these
quotient systems are based on rings (sometimes with additional
spokes as in Fig.~\ref{fig:four}). This explains the synchrony
across the unconnected components of Fig.~\ref{fig:one} -- the
quotient dynamical systems are identical -- a connected pair --
and the dynamics has no phase information as it is just made up
of fixed points. If the dynamics is more complicated then phase
information can be different in different connected components
with the same quotient system, see \cite{bg2011polyad} for more
details.

The emergence of polysynchrony in networks of undifferentiated
nodes operating with a simple homophilic dynamic evolution
suggests natural mechanisms for the emergence of polysynchrony
in nature. Combining this with an additional slow timescale
dynamics to describe the evolution of differentiated
polysynchronous states over time would provide a way of locking
in the polysynchronous pattern making it more stable to
perturbation, or alternatively the evolutionary dynamics could
select the appropriate polysynchronous patterns if only some
subset of the possible patterns is advantageous to the system.
We have reported results with relatively low numbers of nodes.
As the number of nodes increases transient times increase
significantly, and so evolutionary time scales could become
comparable to the time of approach to the frozen state, and
this may provide another way whereby evolution and
differentiation or speciation can interact, although numerical
simulation then becomes harder. The hierarchical nature of the
asymptotic networks imply that they could operate as motifs,
with a well defined input and output level, so in this case our
model may provide a mechanism for functional differentiation.
Models similar to the one proposed here are used to describe
metapopulations \cite{parthasarathy1998synchronisation} so
there may also be applications in this area.

\medskip
\begin{acknowledgments}
PG is partially funded by EPSRC grant EP/E050441/1. VBS is
partially supported by contracts MCyT/FEDER, Spain
FIS2007-60133 and MICINN (AYA2010-22111-C03-02). VBS also
thanks Generalitat Valenciana for financial support.
\end{acknowledgments}


\begin{thebibliography}{40}%
\makeatletter
\providecommand \@ifxundefined [1]{%
 \@ifx{#1\undefined}
}%
\providecommand \@ifnum [1]{%
 \ifnum #1\expandafter \@firstoftwo
 \else \expandafter \@secondoftwo
 \fi
}%
\providecommand \@ifx [1]{%
 \ifx #1\expandafter \@firstoftwo
 \else \expandafter \@secondoftwo
 \fi
}%
\providecommand \natexlab [1]{#1}%
\providecommand \enquote  [1]{``#1''}%
\providecommand \bibnamefont  [1]{#1}%
\providecommand \bibfnamefont [1]{#1}%
\providecommand \citenamefont [1]{#1}%
\providecommand \href@noop [0]{\@secondoftwo}%
\providecommand \href [0]{\begingroup \@sanitize@url \@href}%
\providecommand \@href[1]{\@@startlink{#1}\@@href}%
\providecommand \@@href[1]{\endgroup#1\@@endlink}%
\providecommand \@sanitize@url [0]{\catcode `\\12\catcode
`\$12\catcode
  `\&12\catcode `\#12\catcode `\^12\catcode `\_12\catcode `\%12\relax}%
\providecommand \@@startlink[1]{}%
\providecommand \@@endlink[0]{}%
\providecommand \url  [0]{\begingroup\@sanitize@url \@url }%
\providecommand \@url [1]{\endgroup\@href {#1}{\urlprefix }}%
\providecommand \urlprefix  [0]{URL }%
\providecommand \Eprint [0]{\href }%
\providecommand \doibase [0]{http://dx.doi.org/}%
\providecommand \selectlanguage [0]{\@gobble}%
\providecommand \bibinfo  [0]{\@secondoftwo}%
\providecommand \bibfield  [0]{\@secondoftwo}%
\providecommand \translation [1]{[#1]}%
\providecommand \BibitemOpen [0]{}%
\providecommand \bibitemStop [0]{}%
\providecommand \bibitemNoStop [0]{.\EOS\space}%
\providecommand \EOS [0]{\spacefactor3000\relax}%
\providecommand \BibitemShut  [1]{\csname bibitem#1\endcsname}%
\let\auto@bib@innerbib\@empty
\bibitem [{\citenamefont {Gross}\ and\ \citenamefont
  {Blasius}(2008)}]{Gross2008}%
  \BibitemOpen
  \bibfield  {author} {\bibinfo {author} {\bibfnamefont {T.}~\bibnamefont
  {Gross}}\ and\ \bibinfo {author} {\bibfnamefont {B.}~\bibnamefont
  {Blasius}},\ }\href@noop {} {\bibfield  {journal} {\bibinfo  {journal} {J. R.
  Soc. Interface}\ }\textbf {\bibinfo {volume} {5}},\ \bibinfo {pages} {259}
  (\bibinfo {year} {2008})}\BibitemShut {NoStop}%
\bibitem [{\citenamefont {Gross}\ and\ \citenamefont
  {Sayama}(2009)}]{gross2009adaptive}%
  \BibitemOpen
  \bibfield  {author} {\bibinfo {author} {\bibfnamefont {T.}~\bibnamefont
  {Gross}}\ and\ \bibinfo {author} {\bibfnamefont {H.}~\bibnamefont {Sayama}},\
  }\href@noop {} {\emph {\bibinfo {title} {{Adaptive Networks: Theory, Models
  and Applications}}}}\ (\bibinfo  {publisher} {Springer Verlag},\ \bibinfo
  {year} {2009})\BibitemShut {NoStop}%
\bibitem [{\citenamefont {Boccaletti}\ \emph {et~al.}(2006)\citenamefont
  {Boccaletti}, \citenamefont {Latora}, \citenamefont {Moreno}, \citenamefont
  {Chavez},\ and\ \citenamefont {Hwang}}]{boccaletti2006complex}%
  \BibitemOpen
  \bibfield  {author} {\bibinfo {author} {\bibfnamefont {S.}~\bibnamefont
  {Boccaletti}}, \bibinfo {author} {\bibfnamefont {V.}~\bibnamefont {Latora}},
  \bibinfo {author} {\bibfnamefont {Y.}~\bibnamefont {Moreno}}, \bibinfo
  {author} {\bibfnamefont {M.}~\bibnamefont {Chavez}}, \ and\ \bibinfo {author}
  {\bibfnamefont {D.}~\bibnamefont {Hwang}},\ }\href@noop {} {\bibfield
  {journal} {\bibinfo  {journal} {Phys. Rep.}\ }\textbf {\bibinfo {volume}
  {424}},\ \bibinfo {pages} {175} (\bibinfo {year} {2006})};~
  \BibitemOpen
  \bibfield  {author} {\bibinfo {author} {\bibfnamefont {M.}~\bibnamefont
  {Newman}}, \bibinfo {author} {\bibfnamefont {A.}~\bibnamefont {Barab\'asi}}, \
  and\ \bibinfo {author} {\bibfnamefont {D.}~\bibnamefont {Watts}},\
  }\href@noop {} {\emph {\bibinfo {title} {{The Structure and Dynamics of
  Networks}}}}\ (\bibinfo  {publisher} {Princeton University Press},\ \bibinfo
  {year} {2006})\BibitemShut {NoStop}%
\bibitem [{\citenamefont {Milo}\ \emph {et~al.}(2002)\citenamefont {Milo},
  \citenamefont {Shen-Orr}, \citenamefont {Itzkovitz}, \citenamefont {Kashtan},
  \citenamefont {Chklovskii},\ and\ \citenamefont {Alon}}]{milo2002network}%
  \BibitemOpen
  \bibfield  {author} {\bibinfo {author} {\bibfnamefont {R.}~\bibnamefont
  {Milo}}, \bibinfo {author} {\bibfnamefont {S.}~\bibnamefont {Shen-Orr}},
  \bibinfo {author} {\bibfnamefont {S.}~\bibnamefont {Itzkovitz}}, \bibinfo
  {author} {\bibfnamefont {N.}~\bibnamefont {Kashtan}}, \bibinfo {author}
  {\bibfnamefont {D.}~\bibnamefont {Chklovskii}}, \ and\ \bibinfo {author}
  {\bibfnamefont {U.}~\bibnamefont {Alon}},\ }\href@noop {} {\bibfield
  {journal} {\bibinfo  {journal} {Science}\ }\textbf {\bibinfo {volume}
  {298}},\ \bibinfo {pages} {824} (\bibinfo {year} {2002})};~ 
  \BibitemOpen
  \bibfield  {author} {\bibinfo {author} {\bibfnamefont {U.}~\bibnamefont
  {Alon}},\ }\href@noop {} {\bibfield  {journal} {\bibinfo  {journal} {Nat.
  Rev. Genet.}\ }\textbf {\bibinfo {volume} {8}},\ \bibinfo {pages} {450}
  (\bibinfo {year} {2007})};~ 
  \BibitemOpen
  \bibfield  {author} {\bibinfo {author} {\bibfnamefont {J.}~\bibnamefont
  {Ito}}\ and\ \bibinfo {author} {\bibfnamefont {K.}~\bibnamefont {Kaneko}},\
  }\href@noop {} {\bibfield  {journal} {\bibinfo  {journal} {Phys. Rev. Lett.}\
  }\textbf {\bibinfo {volume} {88}},\ \bibinfo {pages} {028701} (\bibinfo {year}
  {2001})};~
  \BibitemOpen
  \bibfield  {author} {\bibinfo {author} {\bibfnamefont {J.}~\bibnamefont
  {Ito}}\ and\ \bibinfo {author} {\bibfnamefont {K.}~\bibnamefont {Kaneko}},\
  }\href@noop {} {\bibfield  {journal} {\bibinfo  {journal} {Phys. Rev. E}\
  }\textbf {\bibinfo {volume} {67}},\ \bibinfo {pages} {046226} (\bibinfo {year}
  {2003})};~
  \BibitemOpen
  \bibfield  {author} {\bibinfo {author} {\bibfnamefont {Z.}~\bibnamefont
  {Fan}}\ and\ \bibinfo {author} {\bibfnamefont {G.}~\bibnamefont {Chen}},\
  }\href@noop {} {\bibfield  {journal} {\bibinfo  {journal} {Int. J. Mod. Phys.
  B}\ }\textbf {\bibinfo {volume} {18}},\ \bibinfo {pages} {2540} (\bibinfo
  {year} {2004})};~
  \BibitemOpen
  \bibfield  {author} {\bibinfo {author} {\bibfnamefont {D.~V.~D.}\
  \bibnamefont {Berg}}\ and\ \bibinfo {author} {\bibfnamefont {C.~V.}\
  \bibnamefont {Leeuwen}},\ }\href@noop {} {\bibfield  {journal} {\bibinfo
  {journal} {Europhys. Lett.}\ }\textbf {\bibinfo {volume} {65}},\ \bibinfo
  {pages} {459} (\bibinfo {year} {2004})};~
  \BibitemOpen
  \bibfield  {author} {\bibinfo {author} {\bibfnamefont {P.}~\bibnamefont
  {Gong}}\ and\ \bibinfo {author} {\bibfnamefont {C.~V.}\ \bibnamefont
  {Leeuwen}},\ }\href@noop {} {\bibfield  {journal} {\bibinfo  {journal}
  {Europhys. Lett.}\ }\textbf {\bibinfo {volume} {67}},\ \bibinfo {pages} {328}
  (\bibinfo {year} {2004})};~
  \BibitemOpen
  \bibfield  {author} {\bibinfo {author} {\bibfnamefont {C.}~\bibnamefont
  {Zhou}}\ and\ \bibinfo {author} {\bibfnamefont {J.}~\bibnamefont {Kurths}},\
  }\href@noop {} {\bibfield  {journal} {\bibinfo  {journal} {Phys. Rev. Lett.}\
  }\textbf {\bibinfo {volume} {96}},\ \bibinfo {pages} {164102} (\bibinfo {year}
  {2006})};~
  \BibitemOpen
  \bibfield  {author} {\bibinfo {author} {\bibfnamefont {W.}~\bibnamefont
  {Lu}},\ }\href@noop {} {\bibfield  {journal} {\bibinfo  {journal} {Chaos}\
  }\textbf {\bibinfo {volume} {17}},\ \bibinfo {pages} {023122} (\bibinfo
  {year} {2007})};~
  \BibitemOpen
  \bibfield  {author} {\bibinfo {author} {\bibfnamefont {T.}~\bibnamefont
  {Aoki}}\ and\ \bibinfo {author} {\bibfnamefont {T.}~\bibnamefont {Aoyagi}},\
  }\href@noop {} {\bibfield  {journal} {\bibinfo  {journal} {Phys. Rev. Lett.}\
  }\textbf {\bibinfo {volume} {102}},\ \bibinfo {pages} {34101} (\bibinfo
  {year} {2009})};~
  \BibitemOpen
  \bibfield  {author} {\bibinfo {author} {\bibfnamefont {M.}~\bibnamefont
  {Li}}, \bibinfo {author} {\bibfnamefont {S.}~\bibnamefont {Guan}}, \ and\
  \bibinfo {author} {\bibfnamefont {C.-H.}\ \bibnamefont {Lai}},\ }\href@noop
  {} {\bibfield  {journal} {\bibinfo  {journal} {New J. Phys.}\ }\textbf
  {\bibinfo {volume} {12}},\ \bibinfo {pages} {103032} (\bibinfo {year}
  {2010})};~
  \BibitemOpen
  \bibfield  {author} {\bibinfo {author} {\bibfnamefont {A.}~\bibnamefont
  {Scir{\`e}}}, \bibinfo {author} {\bibfnamefont {I.}~\bibnamefont {Tuval}}, \
  and\ \bibinfo {author} {\bibfnamefont {V.}~\bibnamefont {Egu{\'\i}luz}},\
  }\href@noop {} {\bibfield  {journal} {\bibinfo  {journal} {Europhys. Lett.}\
  }\textbf {\bibinfo {volume} {71}},\ \bibinfo {pages} {318} (\bibinfo {year}
  {2005})};~
  \BibitemOpen
  \bibfield  {author} {\bibinfo {author} {\bibfnamefont {T.}~\bibnamefont
  {Gross}}, \bibinfo {author} {\bibfnamefont {C.J.D.}~\bibnamefont {D'Lima}}, \
  and\ \bibinfo {author} {\bibfnamefont {B.}~\bibnamefont {Blasius}},\
  }\href@noop {} {\bibfield  {journal} {\bibinfo  {journal} {Phys. Rev. Lett.}\
  }\textbf {\bibinfo {volume} {96}},\ \bibinfo {pages} {208701} (\bibinfo {year}
  {2006})};~
  \BibitemOpen
  \bibfield  {author} {\bibinfo {author} {\bibfnamefont {T.}~\bibnamefont
  {Gross}}\ and\ \bibinfo {author} {\bibfnamefont {I.}~\bibnamefont
  {Kevrekidis}},\ }\href@noop {} {\bibfield  {journal} {\bibinfo  {journal}
  {Europhys. Lett.}\ }\textbf {\bibinfo {volume} {82}},\ \bibinfo {pages}
  {38004} (\bibinfo {year} {2008})};~
  \BibitemOpen
  \bibfield  {author} {\bibinfo {author} {\bibfnamefont {L.B.}~\bibnamefont
  {Shaw}}\ and\ \bibinfo {author} {\bibfnamefont {I.B.}~\bibnamefont
  {Schwartz}},\ }\href@noop {} {\bibfield  {journal} {\bibinfo  {journal}
  {Phys. Rev. E}\ }\textbf {\bibinfo {volume} {81}},\ \bibinfo {pages} {046120}
  (\bibinfo {year} {2010})};~
  \BibitemOpen
  \bibfield  {author} {\bibinfo {author} {\bibfnamefont {B.}~\bibnamefont
  {Kozma}}\ and\ \bibinfo {author} {\bibfnamefont {A.}~\bibnamefont {Barrat}},\
  }\href@noop {} {\bibfield  {journal} {\bibinfo  {journal} {Phys. Rev. E}\
  }\textbf {\bibinfo {volume} {77}},\ \bibinfo {pages} {016102} (\bibinfo
  {year} {2008})};~
  \BibitemOpen
  \bibfield  {author} {\bibinfo {author} {\bibfnamefont {C.}~\bibnamefont
  {Nardini}}, \bibinfo {author} {\bibfnamefont {B.}~\bibnamefont {Kozma}}, \
  and\ \bibinfo {author} {\bibfnamefont {A.}~\bibnamefont {Barrat}},\
  }\href@noop {} {\bibfield  {journal} {\bibinfo  {journal} {Phys. Rev. Lett.}\
  }\textbf {\bibinfo {volume} {100}},\ \bibinfo {pages} {158701} (\bibinfo
  {year} {2008})};~
  \BibitemOpen
  \bibfield  {author} {\bibinfo {author} {\bibfnamefont {H.}~\bibnamefont
  {Kwok}}, \bibinfo {author} {\bibfnamefont {P.}~\bibnamefont {Jurica}},
  \bibinfo {author} {\bibfnamefont {A.}~\bibnamefont {Raffone}}, \ and\
  \bibinfo {author} {\bibfnamefont {C.}~\bibnamefont {Van~Leeuwen}},\
  }\href@noop {} {\bibfield  {journal} {\bibinfo  {journal} {Cogn. Neurodyn.}\
  }\textbf {\bibinfo {volume} {1}},\ \bibinfo {pages} {39} (\bibinfo {year}
  {2007})};~
  \BibitemOpen
  \bibfield  {author} {\bibinfo {author} {\bibfnamefont {C.}~\bibnamefont
  {Meisel}}\ and\ \bibinfo {author} {\bibfnamefont {T.}~\bibnamefont {Gross}},\
  }\href@noop {} {\bibfield  {journal} {\bibinfo  {journal} {Phys. Rev. E}\
  }\textbf {\bibinfo {volume} {80}},\ \bibinfo {pages} {061917} (\bibinfo
  {year} {2009})};~
  \BibitemOpen
  \bibfield  {author} {\bibinfo {author} {\bibfnamefont {I.}~\bibnamefont
  {Gomez~Portillo}}, \bibinfo {author} {\bibfnamefont {P.}~\bibnamefont
  {Gleiser}}, \ and\ \bibinfo {author} {\bibfnamefont {O.}~\bibnamefont
  {Sporns}},\ }\href@noop {} {\bibfield  {journal} {\bibinfo  {journal} {PLoS
  One}\ }\textbf {\bibinfo {volume} {4}},\ \bibinfo {pages} {418} (\bibinfo
  {year} {2009})};~
  \BibitemOpen
  \bibfield  {author} {\bibinfo {author} {\bibfnamefont {P.}~\bibnamefont
  {Gleiser}}\ and\ \bibinfo {author} {\bibfnamefont {V.}~\bibnamefont
  {Spoormaker}},\ }\href@noop {} {\bibfield  {journal} {\bibinfo  {journal}
  {Philos. T. R. Soc. A}\ }\textbf {\bibinfo {volume} {368}},\ \bibinfo {pages}
  {5633} (\bibinfo {year} {2010})};~
  \BibitemOpen
  \bibfield  {author} {\bibinfo {author} {\bibfnamefont {T.E.}~\bibnamefont
  {Gorochowski}}, \bibinfo {author} {\bibfnamefont {M.}~\bibnamefont
  {di~Bernardo}}, \ and\ \bibinfo {author} {\bibfnamefont {C.S.}~\bibnamefont
  {Grierson}},\ }\href@noop {} {\bibfield  {journal} {\bibinfo  {journal}
  {Phys. Rev. E}\ }\textbf {\bibinfo {volume} {81}},\ \bibinfo {pages} {056212}
  (\bibinfo {year} {2010})};~
  \BibitemOpen
  \bibfield  {author} {\bibinfo {author} {\bibfnamefont {P.}~\bibnamefont
  {DeLellis}}, \bibinfo {author} {\bibfnamefont {M.}~\bibnamefont {di~Bernardo}},
  \bibinfo {author} {\bibfnamefont {T.}~\bibnamefont {Gorochowski}}, \ and\
  \bibinfo {author} {\bibfnamefont {G.}~\bibnamefont {Russo}},\ }\href@noop {}
  {\bibfield  {journal} {\bibinfo  {journal} {IEEE Circuits and Systems}\ }\textbf {\bibinfo {volume} {10}},\ \bibinfo {pages} {64}
  (\bibinfo {year} {2010})}\BibitemShut {NoStop}%
\bibitem [{\citenamefont {Pikovsky}\ \emph {et~al.}(2003)\citenamefont
  {Pikovsky}, \citenamefont {Rosenblum},\ and\ \citenamefont
  {Kurths}}]{pikovsky2003synchronization}%
  \BibitemOpen
  \bibfield  {author} {\bibinfo {author} {\bibfnamefont {A.}~\bibnamefont
  {Pikovsky}}, \bibinfo {author} {\bibfnamefont {M.}~\bibnamefont {Rosenblum}},
  \ and\ \bibinfo {author} {\bibfnamefont {J.}~\bibnamefont {Kurths}},\
  }\href@noop {} {\emph {\bibinfo {title} {{Synchronization: A universal
  concept in nonlinear sciences}}}}\ (\bibinfo  {publisher} {Cambridge Univ.
  Pr.},\ \bibinfo {year} {2003})\BibitemShut {NoStop}%
\bibitem [{\citenamefont {Stewart}\ \emph {et~al.}(2003)\citenamefont
  {Stewart}, \citenamefont {Golubitsky},\ and\ \citenamefont
  {Pivato}}]{stewart2003symmetry}%
  \BibitemOpen
  \bibfield  {author} {\bibinfo {author} {\bibfnamefont {I.}~\bibnamefont
  {Stewart}}, \bibinfo {author} {\bibfnamefont {M.}~\bibnamefont {Golubitsky}},
  \ and\ \bibinfo {author} {\bibfnamefont {M.}~\bibnamefont {Pivato}},\
  }\href@noop {} {\bibfield  {journal} {\bibinfo  {journal} {SIAM J. Appl.
  Dynam. Sys.}\ }\textbf {\bibinfo {volume} {2}},\ \bibinfo {pages} {609}
  (\bibinfo {year} {2003})}\BibitemShut {NoStop}%
\bibitem [{\citenamefont {Golubitsky}\ \emph {et~al.}(2004)\citenamefont
  {Golubitsky}, \citenamefont {Nicol},\ and\ \citenamefont
  {Stewart}}]{golubitsky2004some}%
  \BibitemOpen
  \bibfield  {author} {\bibinfo {author} {\bibfnamefont {M.}~\bibnamefont
  {Golubitsky}}, \bibinfo {author} {\bibfnamefont {M.}~\bibnamefont {Nicol}}, \
  and\ \bibinfo {author} {\bibfnamefont {I.}~\bibnamefont {Stewart}},\
  }\href@noop {} {\bibfield  {journal} {\bibinfo  {journal} {J. Nonlinear
  Sci.}\ }\textbf {\bibinfo {volume} {14}},\ \bibinfo {pages} {207} (\bibinfo
  {year} {2004})}\BibitemShut {NoStop}%
\bibitem [{\citenamefont {Field}(2004)}]{field2004combinatorial}%
  \BibitemOpen
  \bibfield  {author} {\bibinfo {author} {\bibfnamefont {M.}~\bibnamefont
  {Field}},\ }\href@noop {} {\bibfield  {journal} {\bibinfo  {journal}
  {Dynamical Systems}\ }\textbf {\bibinfo {volume} {19}},\ \bibinfo {pages}
  {217} (\bibinfo {year} {2004})}\BibitemShut {NoStop}%
\bibitem [{\citenamefont {Aguiar}\ \emph {et~al.}(2009)\citenamefont {Aguiar},
  \citenamefont {Ashwin}, \citenamefont {Dias},\ and\ \citenamefont
  {Field}}]{aguiar2009dynamics}%
  \BibitemOpen
  \bibfield  {author} {\bibinfo {author} {\bibfnamefont {M.}~\bibnamefont
  {Aguiar}}, \bibinfo {author} {\bibfnamefont {P.}~\bibnamefont {Ashwin}},
  \bibinfo {author} {\bibfnamefont {A.}~\bibnamefont {Dias}}, \ and\ \bibinfo
  {author} {\bibfnamefont {M.}~\bibnamefont {Field}},\ }\href@noop {}
  {\bibfield  {journal} {\bibinfo  {journal} {J. Nonlinear Sci.}\ ,\ \bibinfo
  {pages} {1}} (\bibinfo {year} {2009})}\BibitemShut {NoStop}%
\bibitem [{\citenamefont {Agarwal}\ and\ \citenamefont
  {Field}(2010)}]{agarwal2010dynamical}%
  \BibitemOpen
  \bibfield  {author} {\bibinfo {author} {\bibfnamefont {N.}~\bibnamefont
  {Agarwal}}\ and\ \bibinfo {author} {\bibfnamefont {M.}~\bibnamefont
  {Field}},\ }\href@noop {} {\bibfield  {journal} {\bibinfo  {journal}
  {Nonlinearity}\ }\textbf {\bibinfo {volume} {23}},\ \bibinfo {pages} {1245}
  (\bibinfo {year} {2010})}\BibitemShut {NoStop}%
\bibitem [{\citenamefont {Golubitsky}\ \emph {et~al.}(2005)\citenamefont
  {Golubitsky}, \citenamefont {Stewart},\ and\ \citenamefont
  {T{\"o}r{\"o}k}}]{golubitsky2005patterns}%
  \BibitemOpen
  \bibfield  {author} {\bibinfo {author} {\bibfnamefont {M.}~\bibnamefont
  {Golubitsky}}, \bibinfo {author} {\bibfnamefont {I.}~\bibnamefont {Stewart}},
  \ and\ \bibinfo {author} {\bibfnamefont {A.}~\bibnamefont {T{\"o}r{\"o}k}},\
  }\href@noop {} {\bibfield  {journal} {\bibinfo  {journal} {SIAM J. Appl.
  Dynam. Sys}\ }\textbf {\bibinfo {volume} {4}},\ \bibinfo {pages} {78}
  (\bibinfo {year} {2005})}\BibitemShut {NoStop}%
\bibitem [{\citenamefont {Botella-Soler}\ and\ \citenamefont
  {Glendinning}(2011)}]{bg2011polyad}%
  \BibitemOpen
  \bibfield  {author} {\bibinfo {author} {\bibfnamefont {V.}~\bibnamefont
  {Botella-Soler}}\ and\ \bibinfo {author} {\bibfnamefont {P.}~\bibnamefont
  {Glendinning}},\ }\href@noop {} {\bibfield  {journal} {\bibinfo  {journal}
  {in preparation}\ } (\bibinfo {year} {2011})}\BibitemShut {NoStop}%
\bibitem [{\citenamefont {Quince}\ \emph {et~al.}(2005)\citenamefont {Quince},
  \citenamefont {Higgs},\ and\ \citenamefont {McKane}}]{quince2005topological}%
  \BibitemOpen
  \bibfield  {author} {\bibinfo {author} {\bibfnamefont {C.}~\bibnamefont
  {Quince}}, \bibinfo {author} {\bibfnamefont {P.}~\bibnamefont {Higgs}}, \
  and\ \bibinfo {author} {\bibfnamefont {A.}~\bibnamefont {McKane}},\
  }\href@noop {} {\bibfield  {journal} {\bibinfo  {journal} {Ecol. Model.}\
  }\textbf {\bibinfo {volume} {187}},\ \bibinfo {pages} {389} (\bibinfo {year}
  {2005})}\BibitemShut {NoStop}%
\bibitem [{\citenamefont {Golubitsky}\ \emph {et~al.}(2006)\citenamefont
  {Golubitsky}, \citenamefont {Josi\'c},\ and\ \citenamefont
  {Shea-Brown}}]{golubitsky2006winding}%
  \BibitemOpen
  \bibfield  {author} {\bibinfo {author} {\bibfnamefont {M.}~\bibnamefont
  {Golubitsky}}, \bibinfo {author} {\bibfnamefont {K.}~\bibnamefont {Josi\'c}}, \
  and\ \bibinfo {author} {\bibfnamefont {E.}~\bibnamefont {Shea-Brown}},\
  }\href@noop {} {\bibfield  {journal} {\bibinfo  {journal} {J. Nonlinear
  Sci.}\ }\textbf {\bibinfo {volume} {16}},\ \bibinfo {pages} {201} (\bibinfo
  {year} {2006})}\BibitemShut {NoStop}%
\bibitem [{\citenamefont {Parthasarathy}\ and\ \citenamefont
  {Guemez}(1998)}]{parthasarathy1998synchronisation}%
  \BibitemOpen
  \bibfield  {author} {\bibinfo {author} {\bibfnamefont {S.}~\bibnamefont
  {Parthasarathy}}\ and\ \bibinfo {author} {\bibfnamefont {J.}~\bibnamefont
  {G\"u\'emez}},\ }\href@noop {} {\bibfield  {journal} {\bibinfo  {journal} {Ecol.
  model.}\ }\textbf {\bibinfo {volume} {106}},\ \bibinfo {pages} {17} (\bibinfo
  {year} {1998})};~
  \BibitemOpen
  \bibfield  {author} {\bibinfo {author} {\bibfnamefont {M.}~\bibnamefont
  {Gyllenberg}}, \bibinfo {author} {\bibfnamefont {G.}~\bibnamefont
  {S\"oderbacka}}, \ and\ \bibinfo {author} {\bibfnamefont {S.}~\bibnamefont
  {Ericsson}},\ }\href@noop {} {\bibfield  {journal} {\bibinfo  {journal}
  {Math. Biosci.}\ }\textbf {\bibinfo {volume} {118}},\ \bibinfo {pages} {25}
  (\bibinfo {year} {1993})}\BibitemShut {NoStop}%
\end{thebibliography}
\end{document}